\documentstyle[11pt]{article}
\begin{document}
\thispagestyle{empty}
\rightline{OU-HET 412}
\rightline{SU-ITP-02/20}
\rightline{UOSTP-02103}
\rightline{{\tt hep-th/0205265}}

\makeatletter
\@addtoreset{equation}{section}
\def\theequation{\thesection .\arabic{equation}}
\makeatother

\def\tr{{\rm tr}\,}
\newcommand{\beqn}{\begin{eqnarray}}
\newcommand{\eeqn}{\end{eqnarray}} \newcommand{\bde}{{\bf e}}
\newcommand{\be}{\begin{equation}}
\newcommand{\ee}{\end{equation}}
\pagestyle{plain}
\renewcommand{\baselinestretch}{1.4}
\textwidth 160mm
\textheight 220mm
\newcommand{\bea}{\begin{eqnarray}}
\newcommand{\eea}{\end{eqnarray}}
\newcommand{\vs}[1]{\vspace{#1 mm}}
\newcommand{\hs}[1]{\hspace{#1 mm}}
\renewcommand{\a}{\alpha}
\renewcommand{\b}{\beta}
\renewcommand{\c}{\gamma}
\renewcommand{\d}{\delta}
\newcommand{\e}{\epsilon}
\newcommand{\s}{\sigma}
\def\bbox{{\,\lower0.9pt\vbox{\hrule \hbox{\vrule height 0.2 cm
\hskip 0.2 cm \vrule height 0.2 cm}\hrule}\,}}
\newcommand{\dsl}{\pa \kern-0.5em /}
\newcommand{\la}{\lambda}
\newcommand{\shalf}{\frac{1}{2}}
\newcommand{\pa}{\partial}
\renewcommand{\t}{\theta}
\newcommand{\tb}{{\bar \theta}}
\newcommand{\nn}{\nonumber\\}
\newcommand{\p}[1]{(\ref{#1})}
\newcommand{\lan}{\langle}
\newcommand{\ran}{\rangle}
\newcommand{\NP}[1]{Nucl.\ Phys.\ {\bf #1}}
\newcommand{\PL}[1]{Phys.\ Lett.\ {\bf #1}}
\newcommand{\CQG}[1]{Class.\ Quant.\ Grav.\ {\bf #1}}
\newcommand{\CMP}[1]{Comm.\ Math.\ Phys.\ {\bf #1}}
\newcommand{\IJMP}[1]{Int.\ Jour.\ Mod.\ Phys.\ {\bf #1}}
\newcommand{\JHEP}[1]{J.\ High\ Energy\ Phys.\ {\bf #1}}
\newcommand{\PR}[1]{Phys.\ Rev.\ {\bf #1}}
\newcommand{\PRL}[1]{Phys.\ Rev.\ Lett.\ {\bf #1}}
\newcommand{\PRE}[1]{Phys.\ Rep.\ {\bf #1}}
\newcommand{\PTP}[1]{Prog.\ Theor.\ Phys.\ {\bf #1}}
\newcommand{\PTPS}[1]{Prog.\ Theor.\ Phys.\ Suppl.\ {\bf #1}}
\newcommand{\MPL}[1]{Mod.\ Phys.\ Lett.\ {\bf #1}}
\newcommand{\ket}{\rangle}
\newcommand{\bra}{\langle}

\vskip 1cm
\centerline{
\Large
\bf Supersymmetric Brane-Antibrane Systems:}
\vskip .3cm
\centerline{
\Large
\bf Matrix Model Description, Stability and Decoupling Limits
}

\vskip .2cm

\vskip 1.2cm
\centerline{
Dongsu Bak$^a$, 
 Nobuyoshi Ohta$^{b}$ and Mohammad M. Sheikh-Jabbari$^{c}$
}
\vskip 10mm
\centerline{ \it $^a$ Physics Department,
University of Seoul, Seoul 130-743, Korea}
\vskip 3mm
\centerline{ \it $^b$
Department of Physics, Osaka University,
Toyonaka, Osaka 560-0043, Japan}
\vskip 3mm
\centerline{ \it $^c$
Department of Physics, Stanford University,
Stanford CA 94305-4060, USA
}
\vskip 0.8cm
\centerline{\tt \small 
(dsbak@mach.uos.ac.kr, ohta@phys.sci.osaka-u.ac.jp,
 jabbari@itp.stanford.edu)}

\vskip 1.2cm
\begin{quote}

After reviewing the supertubes and super brane-antibrane systems in the
context of matrix model, we look for more general higher-dimensional
configurations.
For D3-$\overline{\rm D3}$, we find
a non-trivial configuration with ${\bf E}\cdot {\bf B}\neq 0$
and describe the worldvolume gauge theory.
We present the string probe of  D3-$\overline{\rm D3}$ system and study the
decoupling limits leading to either noncommutative
Super-Yang-Mills or NCOS theories with eight supercharges.

\end{quote}

\vskip 3cm
\centerline{\today}
\newpage

\section{Introduction}

After the stringy realization of D-branes~\cite{Polch}, various brane
configurations which preserve some supersymmetry and hence are stable, have
been considered. Among these there are the interesting class of 1/4 BPS
branes, D2-branes of tubular shape generally called supertubes~\cite{mateos}.
The supertubes have also been realized in the matrix theory in
Ref.~\cite{klee},  and it has been shown that it is possible to generalize
the tubular configuration to $p$-branes whose worldvolumes are
$R^{p-1}\times {\cal C}$, where ${\cal C}$ can be a circle or an
ellipse (or even a hyperbola)~\cite{karch}. Later it was proven that
${\cal C}$ can be a completely arbitrary curve~\cite{ng}.

There are simple but intriguing type
of 1/4 BPS brane systems, the supersymmetric brane
anti-branes, which can be thought as a specific limit of the tubular
branes. To see the relation, consider a supertube of elliptic cross section.
In the limit that the longer axis of the ellipse goes
to infinity while
the shorter one is kept fixed, the supertube system becomes the flat
brane-antibrane system~\cite{karch}. The spectrum and stability of the
system were studied in Refs.~\cite{ng,BO} and it was shown that the
brane-antibrane tachyon is stabilized by proper amount of static electric
and magnetic fields on the worldvolume of the branes.
More precisely, for the case of D2-$\overline{\rm D2}$ branes, the
system is stabilized by distributing the proper amount of D0-branes and
F-strings on the worldvolume of D2 and $\overline{\rm D2}$.
Other aspects of supertubes have been discussed in
\cite{{swkim},{CO},{Bena},{mnt},{HO}}.
Related configurations of D2 supertubes ending on D4 have also been studied
through the Born-Infeld action~\cite{{peet}}.

In this paper first we review the matrix model solutions of a $R\times
{\cal C}$ type supertubes as well as 1/4 BPS  D2-$\overline{\rm D2}$ system
where, for the stability of the system, one  needs a very specific electric
field along the $R$ direction (axis of the supertube). However, the
corresponding magnetic field can be chosen arbitrarily (provided that it is
non-zero).
Within the matrix theory,
 we illustrate  how tubes with  arbitrary  cross sectional
shape can be realized.
By a slight modification of the 1/4  BPS equations, we also
construct the solution of string-type spike pulled out of 
D2 with constant B-field background.
Note that the ``source term'' present in this configuration
 is compatible
with the supersymmetry condition.
These solutions can  be easily generalized to the cases of
$p >2$.

In section 3, we focus on the  D3-$\overline{\rm D3}$ system and
consider the worldvolume gauge field configurations 
with ${\bf E}\cdot {\bf B}\neq 0$.
The  supersymmetric D3-$\overline{\rm D3}$ systems, having
less supersymmetries (compared to individual branes) and of higher dimensions,
can be of much interest from phenomenological point of view.
We show that  choosing the electric field of the proper value, i.e. $E=1$
in the conventions of this section, independently of the value of the magnetic
field we can have a 1/4 BPS system. We also analyze this system from
the Born-Infeld point of view. In section 4, starting from the matrix
strings action~\cite{DVV}, we present the matrix theory description of
D3-$\overline{\rm D3}$ system. We also present related BPS equations for
less number of supercharges.

In section 5, we study the spectrum of open strings stretched between D3
and $\overline{\rm D3}$ branes. We show that for a general background
$B_{\mu\nu}$ field there is a tachyon in the spectrum. However, this
tachyon becomes massless in the $E\to 1$ limit, compatible with our arguments
of previous sections.
In section 6, we study the NCSYM limit of the  D3-$\overline{\rm D3}$ brane
system. Performing the Seiberg-Witten limit \cite{seibergwitten}
we show that such a decoupling limit, where the massless open strings are
decoupled from the bulk supergravity modes, exists. In section 7, we show
that the  D3-$\overline{\rm D3}$ brane system admits another ``decoupling
limit'', the noncommutative open string (NCOS) limit. In this limit, unlike
the NCSYM decoupling limit of section 6, we have a non-critical open string
theory, decoupled from all the closed string modes. We show that
the existence of NCOS limit does not depend on the value of the background
magnetic fields (provided that ${\bf E}\times {\bf B}$ is non-zero).
This will provide a new class of NCOS theories with 8 supercharges.
The last section is devoted to conclusions and discussions.

\section{Supertubes in the matrix model}

In this section, we first briefly review how the circular supertubes and super
brane-antibrane systems arise from the matrix model~\cite{klee,swkim}.
We then construct the tubes having an arbitrary cross section within
the matrix model. We further construct the 1/4 BPS configuration
corresponding to (IIA) strings pulled out from noncommutative D2-branes.
This configuration arises in close connection with the tubular geometry
but requires a slight modification of the BPS equation.
In fact this  corresponds to the radius varying tubular configuration
discussed in Ref.~\cite{mateos}.

We begin with the matrix model Lagrangian~\cite{banks1}
\begin{equation}
L={1\over 2 R} \tr \left( \sum_I (D_0 X_I)^2
+{1\over (2\pi \alpha')^2}
\sum_{I<J} [X_I,X_J]^2+ {\rm fermionic\ part}
\right),
\label{lag}
\end{equation}
where $I,J=1,2,\cdots 9$, $R=g_s l_s$ is the radius of the tenth spatial
direction, and $\alpha'(\equiv l_s^2)$ is related to the eleven-dimensional
Planck length by $l_{11}= (R \alpha')^{1\over 3}$.
The scale $R$ together with $2\pi\alpha'$
will be  omitted below by setting them unity.
We would like to emphasize that all our discussions are valid for any
finite $R$ and $\alpha'$ and one does not need any further decoupling
limit for the validity of the description~\cite{seiberg1}.

For the tubular configurations, we note that the supersymmetric variation
of the fermionic coordinates $\psi$ in the matrix theory is
\begin{eqnarray}
\delta \psi =  \left(D_0 X^I\, \gamma_{I} + {i\over 2}[X^I,X^J]
\,\gamma_{IJ}\right)\epsilon +\epsilon'\,,
\label{susy}
\end{eqnarray}
where $\epsilon$ and $\epsilon'$ are real spinors of
16 components parameterizing total 32 supersymmetries.
Here we shall turn on only the first three components of $X^I$, which
will be denoted by $X$, $Y$ and $Z$. Introducing the real projection
operator $P_\pm= (1\pm \gamma_z)/2$, one may rewrite the variation of
the fermionic coordinate as
\begin{eqnarray}
\delta \psi &=&  \left[(D_0 X + i[Z,X])\gamma_x + (D_0 Y +
i[Z,Y])\gamma_y -  D_0 Z + i[X,Y]\gamma_{xy}\right]P_- \e \cr
&+& \left[(D_0 X - i[Z,X])\gamma_x + (D_0 Y - i[Z,Y])\gamma_y +
D_0 Z + i[X,Y]\gamma_{xy}\right]P_+ \e +\e' =0\,.
\label{susya}
\end{eqnarray}
Let us choose $P_+\epsilon =0$ (or $P_-\epsilon=0$) and $\epsilon'=0$ as
remaining
supersymmetries. Any non-trivial solution will then preserve the 1/4 of
the original 32 supersymmetries. The coefficient of $P_-\epsilon$ should
vanish for the remaining supersymmetries; this leads to the BPS equations
\begin{eqnarray}
 D_0 X + i[Z,X]=0\,,\ \
D_0 Y + i[Z,Y]=0\,, \ \  D_0 Z =0\,,\ \ [X,Y]=0\,.
\label{bps1}
\end{eqnarray}
In addition one has to satisfy the Gauss law constraint,
\begin{eqnarray}
[X, D_0 X]+[Y, D_0 Y]+[Z, D_0 Z]=0 \,.
\label{gauss}
\end{eqnarray}
With the choice of the gauge $A_0=Z$, the BPS equations are reduced to
\begin{eqnarray}
 [X,[X,Z]]+ [Y, [Y,Z]]=0\,,\ \ [X,Y]=0\,,
\label{bps2}
\end{eqnarray}
with time independent $X_i$, $i=1,2,3$.\footnote{The time independence
here does not mean that the momentum $D_0X_i$ should vanish.}
The circular tube solution is given by the representation of the algebra,
\begin{eqnarray}
[z,x]=il y, \ \ [y,z]=i l x\,, \ \   [x,y]=0\,,
\label{sol1}
\end{eqnarray}
with $X_i=x_i$.
The shape is described by
the Casimir $\rho^2= x^2+y^2$, which is proportional to identity
for any irreducible
representations of the algebra.
The irreducible representations in the basis
diagonalizing $Z$ are given as~\cite{klee}
\begin{eqnarray}
x+iy=\rho\sum^\infty_{n=-\infty}|n+1\ket\bra n|\,,\ \
z=l\sum^\infty_{n=-\infty}(n+\alpha )|n\ket\bra n|,
\label{sol11}
\end{eqnarray}
with $\alpha\in  [0,1)$. One can check that the configuration has
a circular cross section at any fixed $z$. The location of D0-branes
are equally spaced by $l$ in the $z$-direction having the lattice
translational symmetry. One may generalize the solutions to those of 1/4
BPS multi tubes~\cite{klee}. This implies that there is
no force between these tubes. The momentum density $D_0 X_i$ is conserved
and the central charge measures the F-string
charges~\cite{karch,swkim}.

The elliptic/hyperbolic deformation,
\begin{eqnarray}
[z,x]=ia y, \ \ [y,z]=i b x\,, \ \   [x,y]=0\,,
\label{sol22}
\end{eqnarray}
of the 
algebra also solves the BPS equations for arbitrary constants
$a$ and $b$. With the Casimir operators ${\cal K}= x^2/a +y^2/b$, the shape
is elliptic/hyperbolic if $ab > 0$/$ab <0$. When $ab < 0$, the configuration
is in general composed of two separated sheets for a given ${\cal K}$;
For ${\cal K}> 0$, $a>0$ and $b <0$  the representation is
\begin{eqnarray}
x=\pm \sqrt{a{\cal K}} \,\cosh\kappa\,, \ \
y= \sqrt{|b|{\cal K}}\, \sinh\kappa \,,
\label{sol33}
\end{eqnarray}
with $[\kappa,z]=i\sqrt{|ab|}$.
Each sheet of $\pm$ signature is associated with
an irreducible representation of the algebra.

If one takes the limit where $b\, (>0)$ goes to infinity
while fixing $a \,(>0)$, the cross sectional shape becomes two parallel
lines extended in $y$ direction. Since the total D2 charge of
this system is zero, the configuration in this case corresponds
to brane-antibrane system extended in $y$-$z$ directions, which is 1/4 BPS.

In fact the elliptic/hyperbolic deformation is not the most general case.
The tubes of arbitrary cross sections are possible~\cite{ng}. To see this,
let us consider the configuration defined by
\begin{eqnarray}
X=F_x(x,y)\,,\ \ Y=F_y(x,y)\,,\ \  Z=z,
\label{arbicross}
\end{eqnarray}
where $x,y$ and $z$ are the solutions of (\ref{sol22}) and
$F_x$ and $F_y$ are arbitrary real functions of $x$ and $y$.
One may then easily see that this configuration
solves the BPS equations~\p{bps2}. Since this amounts to an
arbitrary coordinate transformation in $x$-$y$ plane, the curve
defined by $x^2(X,Y)/a+ y^2(X,Y)/b={\cal K}$ takes an arbitrary shape
in $(X,Y)$ space.
To obtain a curve of an arbitrary shape it is enough to introduce one
arbitrary function, say, $F_x$.  The remaining may be associated
with the reparametrization of the coordinate describing the curve for
a given shape. Because of this freedom, the worldvolume $B$-field
become an arbitrary function along the curve but still independent of
$z$. Due to the noncommutative description of the complicated shape,
the argument here  may look unclear. We illustrate  this freedom explicitly
for the case of flat brane solutions.

However before that we would like to
note that there are closely related more general solutions:
\begin{eqnarray}
X=F_x(x,y)\,,\ \ Y=F_y(x,y)\,,\ \  Z={1\over 2}(zH(x,y)+H(x,y)z)+G(x,y),
\end{eqnarray}
where $H$ and $G$ are arbitrary real functions.
This solution is related to (\ref{arbicross}) by a residual
worldvolume reparametrization of the $Z$-coordinate.
The BPS equations in (\ref{bps2}) may be generalized
\begin{eqnarray}
 [X_k,[X_k,X_9]]=0\,,\ \ [X_k,X_l]=0\,,
\label{bps3}
\end{eqnarray}
with $k,l=1,2,\cdots 8$. Then the arbitrary curve may be generalized as
\begin{eqnarray}
X_k=F_k (x,y)\,,\ \  X_9=z,
\label{sol3}
\end{eqnarray}
which describes an arbitrary curve in the eight-dimensional
space transverse to $X_9$.

Now let us consider the planar solutions leading to super brane-antibrane
systems. The configuration is again a particular solution of the above
BPS equations
\begin{eqnarray}
[y,z]=i\Theta \otimes I \,,\ \  x= A \otimes I,
\label{flatsol}
\end{eqnarray}
where $\Theta$ and $A$ are the diagonal $p\times p$
matrices. Here  $p$ is the total number of branes
and the signatures of the diagonal elements $\Theta_{n}$
represent whether the constituent corresponds to a brane or
an antibrane. When $\Theta$ is proportional to identity,
the configuration becomes 1/2 BPS describing $p$ D2-branes
in a constant NS-NS $B$-field background. This enhancement of
supersymmetries may be directly checked using (\ref{susya})
with an appropriate $\epsilon'$.

For the simple brane-antibrane system~\cite{karch},
one may take $p=2$ with $\Theta={\rm diag} (\theta_1,-\theta_2)$ and
$A={\rm diag} (a_1,a_2)$ with $\theta_1,\theta_2 >0$.
The parameter $\theta$ describes the noncommutativity
of the worldvolume theory and $a_1$ and $a_2$ describe
the locations of the brane and antibrane in $x$-direction.
Let us first consider $p=1$ case; the simplest solution is
\begin{eqnarray}
[y,z]=i\theta \,, \ \ x= 0.
\end{eqnarray}
This corresponds to the NS-NS background $B_{23}=-1/\theta$.
It is straightforward to check that the arbitrary
$z$-independent $B$-field, described by
\begin{eqnarray}
Y= y+ \theta h(y)\,,\ \ \ Z=z,
\label{ddbar2}
\end{eqnarray}
where $h$ is an arbitrary real function and $[y,z]=i\theta$, solves the
BPS equations and the shape is still planar. However, if we define gauge
field on the brane as
\[
Y=y+ \theta A_z\,, \ \ Z=z-\theta A_y,
\]
then
\begin{eqnarray}
[Y, Z]=i(\theta + \theta^2 F_{yz})\ .
\end{eqnarray}
Therefore the solution (\ref{ddbar2}) corresponds to  the magnetic field
$F_{yz}=h'(y)$. The background NS-NS $B$-field responsible for the
noncommutativity is $B_{23}=-1/\theta$ with all the other components vanishing.
Thus this planar configuration describes a planar brane with an arbitrary
$B$-field. In this matrix formulation, the worldvolume
dynamics is automatically described by a noncommutative
field theory. The map to commutative variables is not that straightforward.
They are, in fact, related by the Seiberg-Witten map~\cite{seibergwitten};
\begin{eqnarray}
&&A_\mu=a_\mu- {1\over 4}\theta^{\alpha\beta}
\{a_\alpha,\partial_\beta a_\mu +f_{\beta\mu} \} +O(\theta^2), \nonumber\\
&& F_{\mu\nu}= f_{\mu\nu}+{1\over 4}\theta^{\alpha\beta}(
2 \{f_{\mu\alpha},f_{\nu\beta} \}
-\{a_\alpha, D_\beta f_{\mu\nu} +\partial_\beta f_{\mu\nu} \} )
+O(\theta^2),
\label{SW}
\end{eqnarray}
where $\{a,b\}=ab+ba$, the lower case letters are used for the
commutative variables and the capital letter for the noncommutative ones.
The electric field in this solution is evaluated as
$F_{0z} \equiv  D_z A_0- \dot{A}_z=1 + \theta h'(y)$.
Using the Seiberg-Witten map one may confirm that
\begin{eqnarray}
f_{0y}=O(\theta^2)\,,\ \ f_{0z}= 1 +O(\theta^2)\,,\ \
f_{yz}=h'(y)-{\theta\over 2}(h^2(y))'' + O(\theta^2),
\end{eqnarray}
while all the other components are vanishing. Hence we see here that,
to the first order in $\theta$, the electric components agree with the
commutative one.\footnote{It should be noted that only the gauge
invariant combination  $(B + f)_{\mu\nu} $ has a physical meaning.}
Also note that $f_{yz}$ or the combination with NS-NS background
$-1/\theta +f_{yz}$ involve an arbitrary function of $y$,
which is in agreement with the commutative analysis~\cite{ng}.
The agreement is expected to hold to higher orders in $\theta$ but the
confirmation requires the Seiberg-Witten map to higher orders.

Now we discuss another generalization of the BPS equations~\p{bps1}.
Again we turn on the first three component of $X_I$. Using the projection
$P_+\epsilon=0 $ and taking $\epsilon'= \theta\gamma_{12}\epsilon$,
we obtain the following slightly generalized BPS equations:
\begin{eqnarray}
D_0 X + i[Z,X]=0\,,\ \
D_0 Y + i[Z,Y]=0\,,\ \  D_0 Z =0\,,\ \ [X,Y]=i\theta\,.
\label{bps4}
\end{eqnarray}
With the Gauss law and the gauge choice $A_0=Z$, the BPS
equations are reduced to
\begin{eqnarray}
 [X,[X,Z]]+ [Y, [Y,Z]]=0\,,\ \ [X,Y]=i\theta\,.
\label{bps5}
\end{eqnarray}
Without loss of generality, we can assume $\theta> 0$.
Introducing $a=(X+iY)/\sqrt{2\theta}$
and $a^\dagger=(X-iY)/\sqrt{2\theta}$, the second equation
implies $[a, a^\dagger]=1$. In the basis where the number
operator $\hat{n}=a^\dagger a$ is diagonal, the solution reads
\begin{eqnarray}
X-iY = x-iy = \sqrt{2\theta} \sum^\infty_{n=0}
\sqrt{n+1}|n+1\ket\bra n|\ .
\label{plane}
\end{eqnarray}
Observe that the area preserving scaling $\bar{X}= e^\omega X\ $ and
$\bar{Y}= e^{-\omega}Y$ for real $\omega$, generates a new solution.
However, these solutions are related by the following unitary
transformation: $\bar{x}_i=  e^{-\omega (a^{\dagger 2}-a^2)}
x_i e^{\omega (a^{\dagger 2}-a^2)}$. It does not belong to the
gauge transformation as, it does not satisfy the boundary condition at
infinity. Note that the solution in (\ref{plane}) is the one respecting
the  rotational symmetry. Using the solution (\ref{plane}), the
first BPS equation in (\ref{bps5}) becomes the Laplace equation
\begin{eqnarray}
( \partial^2_x + \partial^2_y) Z=0\,,
\label{lap}
\end{eqnarray}
where we define $[x_a, \cdot]=i\theta \epsilon_{ab}\partial_b$
with $a,b=1,2$. In fact using the well known Moyal-Weyl map, $Z$ can be
treated as an ordinary function in $x$-$y$ space. In terms of the
ordinary function $\bar{Z}(x,y)$, Eq.~(\ref{lap})
becomes the usual Laplace equation
and the radially symmetric solution reads
\begin{eqnarray}
 \bar{Z}={Q\over 2\pi}\ln( r/r_0)\,,
\label{radial}
\end{eqnarray}
where we introduced ``source'', $Q \delta^2 (x,y)$,
at $z=-\infty$ (or $r=0$) to the Laplace equation.
Note that the introduction of source term does not modify the
BPS conditions in (\ref{bps4}). The configuration still
preserves 1/4 of the supersymmetries everywhere. The introduction of
source term merely modifies the Gauss law; the source may be consistently
incorporated into the original Lagrangian (\ref{lag}) by a coupling to external
charged sources.

This solution corresponds to a D2-brane with constant $B$-field pulled
out by string. This is precisely the radius varying  solution discussed
in Ref.~\cite{mateos}. Noting $D_0 X_a= -\theta \epsilon_{ab} E_b$,
the electric field evaluated in terms of  functions becomes
$E_a= {Q \over 2\pi} x_a/r^2$ and indeed one can confirm that
the source has the electric charge  $Q$. Furthermore if one uses
$z,\varphi$ coordinates ($\varphi$ is defined by $\tan\varphi=y/x$)
$F_{0z}=\partial_zA_0$ and hence
$E_z=1$, in agreement with Ref.~\cite{mateos}.
The logarithmic behavior comes from the two-dimensional nature of
D2-branes. The presence of point source at the origin makes the D2-brane
get deformed even asymptotically as $r\rightarrow \infty$. This
configuration may straightforwardly be generalized to higher dimensions
in the T-dual setting of the above and then the branes become flat
asymptotically.

The above solutions may be mapped to the operator
representation~\cite{gross} but we shall not delve into this
detailed translation to  the operator representation.
Instead let us study the problem directly solving
the matrix representations. Using the creation annihilation
operator, the BPS equation plus the source  becomes
\begin{eqnarray}
 {1\over \theta}([a,[a^\dagger,Z]]+ [a^\dagger,[a,Z]])=\hat\rho \,,
\label{lap1}
\end{eqnarray}
where $\hat\rho$ is the source.
Although the precise  nature of the source term is
beyond the scope of the classical description, the natural choice from the
view point of the
noncommutative
description would be
\begin{eqnarray}
\hat\rho={Q\over 2\pi\theta} |0\ket\bra 0|  \,,
\label{source}
\end{eqnarray}
which corresponds to the minimal unit of area at the origin.\footnote{
This can be understood as a noncommutative point-like source, as discussed
in~\cite{Susskind}.}
(One may build up general sources
superposing those obtained by
arbitrary translation of
the above  source.) The solution reads
\begin{eqnarray}
Z= \sum^\infty_{n=0}\left(z_0 +
{Q\over 4\pi}\sum^n_{k=1}{1\over k}\right)
|n\ket\bra n| \,.
\label{lapsol}
\end{eqnarray}
Noting that $r^2= 2\hat{n}+1$,  the large $r$ behavior agrees with
(\ref{radial}). For small $n$, the matrix solution (\ref{lapsol}) and
the functional solution (\ref{radial}) do not agree because
$|0\ket\bra 0|$ corresponds  not to a delta function source  but a
Gaussian type, in the function description.


\section{$\kappa$-symmetry of supersymmetric D3-$\overline{\rm\bf D3}$}

In the D2-$\overline{\rm D2}$ system discussed
in the previous section, the $B$-field components turned on
are $B_{23}$ and $E_z= B_{30}$. As mentioned earlier,
$E_z=1$ while $B_{23}$ may be an arbitrary function of $y$. The obvious 
higher-dimensional generalizations  may be obtained by the T-duality 
actions along the transverse directions, e.g. in this way we can obtain a 
D3-$\overline{\rm D3}$ system but with
$E$ and $B$ fields satisfying ${\bf E} \cdot {\bf B}=0$.

Here we ask for more general configurations.
In particular we consider turning on additional component of $B$ in the 
$z$-direction
in D3-$\overline{\rm D3}$ system.
This will lead to the configuration of  ${\bf E} \cdot {\bf B}\neq 0$.
The supersymmetry of the configuration is determined by
\begin{eqnarray}
\Gamma \epsilon_L = \pm \epsilon_R,
\label{susy1}
\end{eqnarray}
where $\Gamma$ is the projection operator appearing in the worldvolume
$\kappa$ symmetry and $+/-$ signs are respectively for brane/antibrane.
The projection operator is determined as~\cite{bergshoeff}
\begin{eqnarray}
\Gamma ={\sqrt{g}\over\sqrt{g+B}}
(\gamma_{0123}-\gamma_{12}E + \gamma_{03} \tilde{B}
+\gamma_{01}  B -E  \tilde{B})\,,
\end{eqnarray}
where $E=B_{03}$, $B=B_{23}$ and $\tilde{B}=B_{12}$.
Then the condition (\ref{susy1}) may be rearranged as
\begin{eqnarray}
{\sqrt{g}\over\sqrt{g+B}}
[(\gamma_{12} +\tilde{B})(\gamma_{03}-E)
+\gamma_{01}  B] \epsilon_L=\pm \epsilon_R
\,.
\label{kappa}
\end{eqnarray}
When $E$, $B$ and $\tilde{B}$ vanish, the supersymmetry condition
reduces to that of D3 (or $\overline{\rm D3}$)-branes, i.e.
$\gamma_{0123} \epsilon_L=\pm \epsilon_R$. Without $B$-fields,
it is clear that there is no remaining supersymmetry
if both brane and antibrane are present.
As in the case of supersymmetric D2-$\overline{\rm D2}$
system, we set $E=1$ with projection $P_+^z\epsilon_L=\epsilon_L$
where we have defined
\begin{eqnarray}
P_\pm^z\equiv  (1\pm \gamma_{03})/2\,.
\end{eqnarray}
Then the remaining supersymmetry condition becomes\footnote{Note that
here we have set $g_{\mu\nu}=\eta_{\mu\nu}$.}
\begin{eqnarray}
{\rm sign}(B)\gamma_{01}  \epsilon_L=\pm \epsilon_R
\,.
\label{bpscon}
\end{eqnarray}
This may be solved for both signs with the same $\epsilon_R$
if $B > 0$ for $+$ sign and $B<0$ for $-$ sign.
As in the case of supersymmetric D2-$\overline{\rm D2}$, only the
signature of $B$ is correlated to the brane charges and
the magnitude is arbitrary. In addition $\tilde{B}$ is completely
arbitrary and could change from brane to brane; the supersymmetries
are not affected by them. The resulting configurations preserve 1/4 of 32
supersymmetries.
The supersymmetry condition by the above projection is
that of stretched strings in $z$-direction and
the condition in (\ref{bpscon}) agrees with supersymmetry condition for
D-strings stretched in $x$-direction. This indicates that the system
may be interpreted as a bound state of branes composed of F-strings and
D-strings. The presence of $\tilde{B}$ implies that there are also
D-strings stretched along $z$ direction inside the D3-branes.

In the above we have considered a constant $B$ and $E$ fields allowing
variations from brane to brane.
Let us now look more closely at the Born-Infeld
description of such branes. The action is
\begin{eqnarray}
{\cal L}=- \sqrt{-\det (g+B)}=-\sqrt{1-{\bf E}^2+{\bf B}^2-
({\bf E}\cdot{\bf B})^2}
=-\sqrt{(1-E^2)(1+\tilde{B}^2)+B^2},
\label{action}
\end{eqnarray}
where we take $g_{\mu\nu}=\eta_{\mu\nu}$ for simplicity.
However as we will see, it would be more convenient to consider
more general closed string metric when we discuss the Seiberg-Witten
decoupling, as well as the NCOS limit.
The displacement becomes
\begin{eqnarray}
{\bf \Pi}={\partial{\cal L}\over\partial  {\bf E} }=
{{\bf E}+{\bf B}\, ({\bf E}\cdot {\bf B})\over
\sqrt{1-{\bf E}^2+{\bf B}^2-
({\bf E}\cdot{\bf B})^2}}
\label{momentum}
\end{eqnarray}
and the Hamiltonian is given by
\begin{eqnarray}
{\cal H}=
\sqrt{1+{\bf B}^2+{\bf \Pi}^2 +({\bf \Pi}\times {\bf B})^2}
\end{eqnarray}
This may be arranged to the complete squared form as
\begin{eqnarray}
{\cal H}=
\sqrt{\Bigl(\sqrt{1+\tilde{B}^2}-{\Pi |B|\over \sqrt{1+\tilde{B}^2}}
\Bigr)^2 +
\Bigl(\tilde{\Pi}\sqrt{1+\tilde{B}^2}-
{\Pi B \tilde{B}\over \sqrt{1+\tilde{B}^2}}
\Bigr)^2
+(|B|+\Pi)^2} \ \ \ge |B|+\Pi
\,,
\end{eqnarray}
where $\Pi= \Pi_z$ and $\tilde{\Pi}=\Pi_x$. This analysis leads to
the BPS equations
\begin{eqnarray}
|B|\Pi=1+\tilde{B}^2
\,,\ \ \  \tilde{\Pi}(1+\tilde{B}^2) =\Pi B \tilde{B}\,,
\end{eqnarray}
which implies $E^2=1$.

The Bianchi identity $d F=0$  and the Gauss law constraint $\partial_z\Pi+
\partial_x \tilde{\Pi}=0$
lead to the conditions
\begin{eqnarray}
\partial_x B+\partial_z \tilde{B}=0\,,\ \ \
\partial_z \left({1+\tilde{B}^2\over |B|}\right)+
\partial_x \left(\tilde{B}\, {B\over |B|}\right)=0\,.
\label{bianchi}
\end{eqnarray}
With these, one may check explicitly that all
the equations of motion are satisfied.

Assuming $\partial_z B= \partial_z \tilde{B}=0$,
the above conditions imply that
$B$ and $\tilde{B}$ are  arbitrary functions of $y$ variable only,
i.e.
\begin{eqnarray}
B= B(y)\,,\ \  \tilde{B}= \tilde{B}(y)\,,\ \ E=1\,.
\label{arbi}
\end{eqnarray}
These 1/4 BPS configurations have a translational symmetry in $z$ and
$x$-direction.

We do not know how to solve (\ref{bianchi}) in general and any nontrivial
solution depending on $x$ or $z$ would be quite interesting. A particular
example of the solution with $\partial_z B\neq 0$ may be given by
$B=C x \sec^2 C z$ and $\tilde{B}= -\tan Cz$, which are singular at
$Cz= n\pi/2$ with odd integer $n$.


\section{Matrix model description of D3-$\overline{\rm{\bf D3}}$ system}

In this section, we will try to reconstruct the D3-$\overline{\rm D3}$
using the matrix model. Specifically we shall use the matrix model
compactified on a circle, matrix strings~\cite{DVV}, by which one may
obtain branes in IIB theory compactified on the dual circle.
The bosonic part of the Lagrangian is~\cite{DVV}
\begin{equation}
L={1\over 2 }\int d\sigma
\tr \left( ({\cal F}_{t\sigma})^2+
\sum_I (D_0 X_I)^2 -(D_\sigma X_I)^2
+
\sum_{I<J} [X_I,X_J]^2+ {\rm fermionic\ part} \right),
\label{lagn}
\end{equation}
where $I,J=2, \cdots, 9$ and the matrix $X_I$ are function of $(t,\sigma)$.
{}From the original matrix model, the $X^1$-direction is compactified on
the circle and hence after the T-duality it is replaced with the gauge
field $A_{\sigma}$ in the above Lagrangian and we consider the theory
in the decompactification limit of the large dual circle.

The BPS equation related to the previous discussion may be
obtained from the variation of the fermionic coordinates.
Alternatively one may obtain the BPS equations by studying
the Hamiltonian. Here we follow the latter method.
Consider again turning on only $X,Y,Z$ of the transverse
components. The Hamiltonian can be arranged as
\begin{eqnarray}
&&H={1\over 2 } \tr \left( ({\cal F}_{t\sigma}- D_\sigma Z)^2+
(D_0 X+ i[Z,X])^2+(D_0 Y + i[Z,Y])^2\right.\nonumber\\
&&\ \ \ \ \ \left. +(D_0 Z)^2 + (D_\sigma X)^2+ (D_\sigma Y)^2
+|[X,Y]|^2+ 2 C_J
\right)\  \ge\   \tr C_J,
\label{hamiltonian}
\end{eqnarray}
where the central charge $\tr C_J$ is defined by
\begin{equation}
\tr C_J= i\,\tr\Bigl( [X, Z (D_0 X)]+[Y, Z (D_0 Y)] +\partial_\sigma
({\cal F}_{t\sigma} Z)\Bigr)\,.
\label{centralcharge}
\end{equation}
Note that to obtain (\ref{centralcharge}) we have used the Gauss law
constraint,
\begin{equation}
-i[X_I, D_0 X_I] +D_\sigma
{\cal F}_{t\sigma} =0\,.
\label{gaussn}
\end{equation}
Then the BPS equations are
\begin{eqnarray}
&& {\cal F}_{t\sigma}- D_\sigma Z=0\,,\ \ D_0 X+ i[Z,X]=0\,,\ \
D_0 Y + i[Z,Y]=0, \nonumber\\
&& [X,Y]=0\,,\ \ \ \  D_0 Z=D_\sigma X=D_\sigma Y=0.
\label{bps6}
\end{eqnarray}
Choosing the gauge $A_0=Z$, the BPS equations imply that
all the matrix variables should be static. Together with
the Gauss law~\p{gaussn}, the BPS equations~\p{bps6} are reduced to
\begin{eqnarray}
&& [X,[X,Z]]+[Y,[Y,Z]]+ D^2_\sigma Z=0\,,\nonumber\\
&& [X,Y]=0\,, \ \ \ \ D_\sigma X=D_\sigma Y=0.
\label{bpsnn}
\end{eqnarray}

To obtain the BPS equations one could use the remaining supersymmetry
condition using the variation of the fermionic coordinate. The 1/4 BPS
condition we are interested in is the T-dual version of the tubular branes.
Note that there we have used the projection $P_\pm=(1\pm \gamma_z)/2$.
While keeping this projection, the condition ${\cal F}_{t\sigma}= D_\sigma Z$
is inevitable for the 1/4 BPS configuration.

Before solving these BPS equations, we would like to record here more
general form of the BPS equations by turning on the remaining transverse
coordinates. By a similar method, it is straightforward to show
that the resulting BPS equations are
\begin{eqnarray}
&& \sum^8_{k=2}[X_k,[X_k, X_9]]+ D^2_\sigma  X_9=0\,,\nonumber\\
&& [X_k,X_l]=0 \,, \ \ \ \ D_\sigma X_k=0,
\label{IIB}
\end{eqnarray}
where $k,l=2,3,\cdots 8$.

The T-dual version of the
above BPS equations are already given in (\ref{bps3}).
Further generalization analogous to \p{bps5} is given by
\begin{eqnarray}
\sum^8_{k=1}[X_k,[X_k, X_9]]=0\,,\ \ \ 
[X_k,X_l]=i\theta_{kl} \,,
\end{eqnarray}
where $k,l=1,2\cdots 8$. There are solutions of this BPS equation
generalizing the IIA D2$p$-branes in a constant NS-NS $B$ field background
pulled out by fundamental strings. We shall not repeat here the previous
discussion. Similar configuration can be found in the IIB theory
generalizing (\ref{IIB}).

The above sets of BPS equations are direct generalizations of (\ref{bps2})
and (\ref{bps5}) to higher dimensions. Their solutions preserve at least a
quarter of the original 32 supersymmetries. There are other generalizations
involving smaller fractions. For example, the BPS equations for 1/8
supersymmetries may be constructed as follows. We first introduce
projection operators
\begin{eqnarray}
 P^9_\pm={1\pm\gamma_9\over 2}\,,\ \ \
P^{1234}_\pm={1\pm\gamma_{1234}\over 2}\,,
\end{eqnarray}
which are real and commute with each other.
Using the variation of the fermionic coordinate in (\ref{susy}),
we consider the cases with
\begin{eqnarray}
P^9_- \epsilon &=& 0\ , \ \  \ \ P^{1234}_- \epsilon = 0 \cr
\epsilon' &=& {1\over 2}{\theta^{mn}\gamma_{mn}}\epsilon\ ,
\ \ \  m,n,p,q=1,2,3,4\ .
\end{eqnarray}
This leads to the BPS equations
\begin{eqnarray}
D_0 X^m=-i[X^9,X^m]\,,\ \ D_0 X^9=0\,,\ \
[X^m,X^n]-i\theta^{mn}=
{1\over 2}\epsilon_{mnpq}([X^p,X^q]-i\theta^{pq})
\end{eqnarray}
with the Gauss constraint
\begin{eqnarray}
[X^m,D_0 X^m]=0\,.
\end{eqnarray}
In the gauge where $A_0=X^9$, these equations are reduced to
\begin{eqnarray}
&&[X^m,X^n]-i\theta^{mn}=
{1\over 2}\epsilon_{mnpq}([X^p,X^q]-i\theta^{pq})\nonumber\\
&&[X^m,[X^m,X^9]]=0,
\label{1/8bps}
\end{eqnarray}
with all the variables independent of time.
The first equation here is for  noncommutative instantons of D0-D4
systems and the second makes them dyonic. These are 1/8 BPS
equations for the noncommutative version of the dyonic instantons
\cite{tong} and the supertubes ending on D4-brane \cite{peet}.
Detailed investigations for $U(1)$ as well as $U(p)$ would be quite
interesting but we shall not attempt here to do so.

Now we get back to the BPS equations in (\ref{bpsnn}). The solutions
\begin{eqnarray}
 X=0\,,\ \  Y=y+ \theta h(y)\,,\ \   Z=z\,,\ \ A_\sigma=0,
\label{ddbar3}
\end{eqnarray}
with $[y,z]=i\theta$ correspond to the T-dual version
of the flat D2 solution in (\ref{ddbar2}). We  shall identify
the worldvolume directions as $x^1=\sigma$, $x^2=y$ and
$x^3=z$. $X$ is then describing the transverse
fluctuation of the D3-brane. Again using the definition
\begin{eqnarray}
Y=x^2+ \theta A_3\,,\ \   Z=x^3-\theta A_2\,,
\end{eqnarray}
one finds that
\begin{eqnarray}
[Y,Z]=i\theta +i\theta^2 F_{23}\,,\ \
D_a Z=\theta F_{a2}\,,\ \
D_a Y=-\theta F_{a3}\,,\ \ F_{t\sigma}=F_{01},
\end{eqnarray}
with $a=0,1$. The solution (\ref{ddbar3}) is a special case of above with
$F_{12}=0$ and $F_{23}= h'(y)$ and only the $B_{23}$ component of NS-NS
two-form background is non-vanishing. Moreover, using the Seiberg-Witten
map, one can show that $f_{03}=1$ and $f_{23}$ is an arbitrary function
of $y$. Hence it is clear that the above solution
is the T-dual version of (\ref{ddbar2}) in the $\sigma$ direction.

Let us now consider the case with non-zero $F_{12}$.
There is a simple solution with non-vanishing $\tilde{B}$;
\begin{eqnarray}
 X=0\,,\ \  Y=y+ \theta h(y)\,,\ \   Z=z\,,\ \ A_\sigma=k(y),
\label{aaa}
\end{eqnarray}
with $[y,z]=i\theta$.
The solution stands for a flat brane and $h(y)$ and $k(y)$ are completely
arbitrary functions of $y$. Of course there are more general solutions
which depend on $x^1=\sigma$ but we shall restrict our consideration to
the $\sigma$ independent cases. For the solution (\ref{aaa}), $F_{yz}=h'(y)$,
$F_{12}= k'(y)$ and $F_{t\sigma}= -\theta k'$.
At first sight, the comparison to that of the Born-Infeld
analysis may seem problematic. Namely there is non-vanishing
$F_{01}$ component (related to $F_{12}$), which was not present
in the analysis of the previous section.
However again the Seiberg-Witten map plays a non-trivial role.
For the comparison, we again use the Seiberg-Witten map, Eqs.~(\ref{SW})
to get the commutative variables. They are evaluated as
\begin{eqnarray}
&& f_{01}=f_{02}=f_{13}=O(\theta^2)\,,\ \
f_{03}= 1 +O(\theta^2)\,,\nonumber\\
&& f_{12}= k'(y)-{\theta}(hk')' + O(\theta^2)\,,\ \
f_{23}=h'(y)-{\theta\over 2}(h^2(y))'' + O(\theta^2).
\label{222}
\end{eqnarray}
Hence these are in agreement with the analysis of the
previous section.

Many brane solutions generalizing (\ref{aaa})
can be easily obtained in a similar way described in the previous section.
Here we shall not exhaust all possible solutions of the above
BPS equations.  There are many obvious solutions. There exist solutions
corresponding to the three-dimensional generalization of tubes of
arbitrary cross section obtained by T-duality.


Finally let us consider the worldvolume dynamics
around the supersymmetric background,
\begin{eqnarray}
X=0\,,\ \  Y=y\,,\ \   Z=z\,,\ \ A_\sigma=\tilde{B} y
\label{aaa1}
\end{eqnarray}
with constant  $\tilde{B}$ and $[y,z]=i\theta$. This corresponds to
$h=0$ and $k=\tilde{B}y$ of the solution (\ref{aaa}).
The non-vanishing components of NS-NS two-form background is then
$B_{23}=-1/\theta$, $F_{12}=\tilde{B}$ and $F_{01}=-\theta \tilde{B}$.
Through the computation of the open string metric on the worldvolume,
we demonstrate that this solution precisely describes the supersymmetric
background in (\ref{arbi}) with constant $B=-1/\theta$ and $\tilde{B}$.
This identification is also consistent with the results in (\ref{222}).
To describe the worldvolume dynamics, let us define the worldvolume
gauge fields ${\cal A}_\mu$ by
\begin{eqnarray}
&& A_0= Z+{\cal A}_0\,,\ \
A_\sigma = \tilde{B} Y+{\cal A}_1\nonumber\\
&& Y=y+\theta {\cal A}_3\,,\ \  Z=z-\theta {\cal A}_2\,.
\end{eqnarray}
Also $X_s=\varphi_s$, where $X_s$ are other than $Y,Z$, become the six
transverse scalar fields.
Inserting these  to the matrix model Lagrangian (\ref{lagn}), one finds
\begin{eqnarray}
\hs{-10}
L&\!\! = \!\!&{1\over 2}\int d\sigma \tr \left[ {\cal F}^2_{01}+
\theta^2 {\cal F}^2_{02}  +\theta^2 (1\!+\!\tilde{B}^2)
{\cal F}^2_{03}\! -\!\theta^2 {\cal F}^2_{13}
+2\theta ( {\cal F}_{01}\!+\!\theta\tilde{B} {\cal F}_{03})
({\cal F}_{12}\!+\!\theta\tilde{B} {\cal F}_{32} )
\right.\nonumber\\
\hs{-10}
&\!\! + \!\!&\left.
2\theta  {\cal F}_{03} (\tilde{B}{\cal F}_{01}\!-\!\theta
 {\cal F}_{23})
\!+\!(\nabla_0\varphi_s\!-\!\theta\nabla_2\varphi_s)^2
\!-\!(\nabla_1\varphi_s\!+\!\theta\tilde{B}\nabla_3\varphi_s)^2
\!-\!\theta^2
(\nabla_2\varphi_s)^2\!-\!\theta^2(\nabla_3\varphi_s)^2 \right]\!\,,
\label{world1}
\end{eqnarray}
where
\begin{equation}
\nabla_\mu=\partial_\mu-i[{\cal A}_\mu,\ ]\,,\ \ \
{\cal F}_{\mu\nu}=\partial_\mu {\cal A}_\nu-
\partial_\nu {\cal A}_\mu-i[{\cal A}_\mu,{\cal A}_\nu]\,.
\end{equation}
This Lagrangian may be written in a standard form
\begin{eqnarray}
&&L=-{1\over 4 g^2_{\rm YM}}\int d\sigma 2\pi\theta\tr \sqrt{G}\left(
G^{\alpha\mu}G^{\beta\nu} {\cal F}_{\alpha\beta} {\cal F}_{\mu\nu}
+ 2 G^{\mu\nu}\nabla_\mu\varphi_s \nabla_\nu\varphi_s
\right)
\end{eqnarray}
with help of an appropriate metric $G^{\mu\nu}$.
In section 6, we shall explicitly show that
the metric appearing here is precisely the open string metric
associated with the background (\ref{arbi}).

\section{D3-$\overline{\rm \bf D3}$ open string spectrum and stability}

As we have shown explicitly in section 3, we can have a ${1\over 4}$ BPS
D3-$\overline{\rm D3}$ brane system with $B$-fields turned on them.
In particular we discussed that  there is a $B$-field configuration with
${\bf E}\cdot {\bf B}\neq 0$. These
solutions are described by non-zero $B_{03}, B_{12}, B_{23}$ fields.
Such D3- and $\overline{\rm 
D3}$-branes can be obtained by T-duality in a tilted 
direction on the solutions of Ref.~\cite{karch,BO}, therefore we expect that
there should not be a tachyonic mode in the open string spectrum stretched 
between the D3- and $\overline{\rm D3}$-branes.
For the later use and as an  explicit check, 
in this section we work out the details of these 
open string spectrum.

The background fields at $\s=0$ (on the D3-brane) are
\bea\label{BBB}
B^{(0)} = \frac{1}{2\pi \a'} \left(
\begin{array}{cccc}
0 & 0 & 0  & E \\
0 & 0 & \tilde{B} & 0 \\
0 & -\tilde{B} & 0 & B \\
-E & 0 & -B & 0
\end{array}
\right); \quad
g_{ij} = \eta_{ij},
\label{back0}
\eea
and as we have discussed earlier, at $\s=\pi$ (on the $\overline{\rm D3}$-brane)
$B^{(\pi)}$ is obtained by reversing the sign of $B$ and $\tilde{B}$.
The boundary conditions are given as
\bea
g_{ij} \pa_\sigma X^j + 2\pi \a' B_{ij} \pa_t X^j =0,\ \ \sigma=0,\pi\ .
\eea
In a more explicit form they are
\bea
\pa_\s X^0 - E \pa_t X^3 &=& 0, \nn
\label{bc1}
\pa_\s X^1 \pm \tilde{B}
\pa_t X^2 &=& 0, \quad {\rm at} \;\; \sigma=0,\pi, \\
\pa_\s X^2 \mp \tilde{B} \pa_t X^1 \pm B \pa_t X^3 &=& 0, \nn
\pa_\s X^3 - E \pa_t X^0 \mp B \pa_t X^2 &=& 0. \nonumber
\eea
The mode expansions are found to be
\bea
X^0 &=& \frac{1}{\sqrt{1-E^2}}(x^0 + 2\a' p_0 t)
 + \frac{E\tilde{B}}{B} 2\a' p_1 \s \nn
&& + i \sqrt{\frac{2\a'}{1-E^2}} \sum_{n\neq 0} \left[
 \sqrt{1+\frac{\tilde B^2 E^2}{\rho^2}} \frac{a_{n}^0}{n} e^{-int} \cos n\s
 - \frac{E\tilde{B}}{\rho} \frac{a_n^1}{n} e^{-in(t-\s)} \right]\nn
&& \hs{-5} - \; i\frac{EB\sqrt{\a'}}{(1-E^2)\rho} \sum \left(
 \frac{c_{n+\nu}}{n+\nu} e^{-i(n+\nu)t - i\frac{\pi\nu}{2}}
- \frac{d_{-n-\nu}}{n+\nu} e^{i(n+\nu)t +i\frac{\pi\nu}{2}}\right)
\sin\left[(n+\nu)\s -\frac{\pi\nu}{2}\right], \nn
\label{Xs}
X^1 &=& x^1 +2\a' p_1 t + i \sqrt{\frac{2\a'}{1-E^2}}\frac{B}{\rho}
\sum_{n\neq 0} \frac{a_{n}^1}{n} e^{-int} \cos n\s \nn
&& - \; \frac{\tilde{B}\sqrt{\a'}}{\rho} \sum \left(
 \frac{c_{n+\nu}}{n+\nu} e^{-i(n+\nu)t - i\frac{\pi\nu}{2}}
+ \frac{d_{-n-\nu}}{n+\nu} e^{i(n+\nu)t +i\frac{\pi\nu}{2}}\right)
\cos\left[(n+\nu)\s -\frac{\pi\nu}{2}\right], \nn
X^2 &=& x^2+ i \sqrt{\a'} \; \sum \left(
 \frac{c_{n+\nu}}{n+\nu} e^{-i(n+\nu)t - i\frac{\pi\nu}{2}}
- \frac{d_{-n-\nu}}{n+\nu} e^{i(n+\nu)t +i\frac{\pi\nu}{2}}\right)
\cos\left[(n+\nu)\s -\frac{\pi\nu}{2}\right], \\
X^3 &=& x^3+\frac{\tilde{B}}{B} 2\a' p_1 t
 + \frac{E}{\sqrt{1-E^2}}2\a' p_0 \s \nn
&& + \sqrt{\frac{2\a'}{1-E^2}} \sum_{n\neq 0} \left[
 E \sqrt{1+\frac{\tilde B^2 E^2}{\rho^2}} \frac{a_n^0}{n} \sin n\s
+ i \frac{\tilde{B}}{\rho}(\cos n\s + iE^2 \sin n\s) \frac{a_n^1}{n}\right]
 e^{-int} \nn
&& + \; \frac{B\sqrt{\a'}}{(1-E^2)\rho} \sum \left(
 \frac{c_{n+\nu}}{n+\nu} e^{-i(n+\nu)t - i\frac{\pi\nu}{2}}
+ \frac{d_{-n-\nu}}{n+\nu} e^{i(n+\nu)t +i\frac{\pi\nu}{2}}\right)
\cos\left[(n+\nu)\s -\frac{\pi\nu}{2}\right], \nonumber
\eea
where $\nu$ and $\rho$ are defined by
\bea
\tan\frac{\pi\nu}{2} = \sqrt{ \tilde{B}^2 +
\frac{B^2}{1-E^2}} \; \equiv \rho,\ \ \ \
 ( 0 \le  \nu < 1),
\eea
and the commutation relations of the mode operators are
\bea
[c_{n+\nu}, d_{m-\nu} ] = (n+\nu) \d_{n+m}, \quad
[a^0_n, a^0_m] = -n \d_{n+m}, \quad
[a^1_n, a^1_m] = n \d_{n+m}, \quad
[x^\mu, p_\nu] = i\eta^\mu{}_\nu.
\eea
This result shows that $\nu=1$ for $E=1$, in which case the
levels are integer, which is in accordance with the supersymmetry
restoration in the limit $E=1$.\footnote{Though the mode expansions~\p{Xs} look
singular in the limit $E \to 1$, the worldsheet CFT is well defined
just as for D2-$\overline{\rm D2}$ system~\cite{BO}.}

We define vacuum by $c_{n+\nu}|0\rangle =0$ for $n \ge 0$
and $d_{n-\nu} |0\rangle =0 $ for $n >0$. Hence the corresponding vacuum
energy from $X^0, X^1, X^2$ and $X^3$ becomes
\bea
E_\nu = -{1\over 2}\sum^\infty_{n=1} n +{1\over 2}\sum^\infty_{n=1} n
+{1\over 2}\sum^\infty_{n=1} (n-\nu) +
{1\over 2}\sum^\infty_{n=1} (n-(1-\nu))
= {1\over 24}-{1\over 8}(2\nu-1)^2 \,,
\eea
where we have used
\bea
{\cal E}(a) \equiv \sum^\infty_{n=1} (n-a) =
{1\over 24}-{1\over 8}(2a-1)^2\,.
\eea

The vacuum energy for the Ramond sector is trivial; the bosonic one is
canceled by the fermionic contribution and the total is zero.
In the Neveu-Schwarz sector, the vacuum energy from the NS-fermions
associated with $0,1,2,3$ components is given by the substitution
$\nu \to |\nu-1/2|$. In the ``light-cone'' gauge, the contributions from
$a_n^0$ and one transverse oscillator cancel with each other. Summing the
contributions from the rest of oscillators, we find the ground state energy
\bea
E_\nu^{total} &=& \left( {1\over 24}-{1\over 8}(2\nu-1)^2\right)
 - \left( {1\over 24}-{1\over 8}(2|\nu-1/2|-1)^2\right)
-\frac{6}{24} - \frac{6}{48} \nn
&=&-{1\over 4} -\frac{|\nu -1/2|}{2} \,.
\label{g}
\eea
For $\nu=0$, this is consistent with the vacuum energy of
NS sector when the background $B$-fields are turned off.

The ground state energy~\p{g} gives two states with energy
\bea
E_1 = -\frac{1}{2} \nu, \;\; &{\rm for}& \;\; 1 \geq \nu \geq \frac{1}{2}, \nn
E_0 = \frac{1}{2} (\nu-1) , \;\; &{\rm for}& \;\; \frac{1}{2} \geq \nu \geq 0.
\eea
When $\nu=0$, $|E_0\ran$ gives the true ground state and $|E_1\ran$ is
the first excited state, but the energy changes when $\nu$ is increased.
For $\nu \geq \frac{1}{2}$, $|E_1\ran$ becomes the true ground state and
$|E_0\ran$ is the first excited state.
For $\nu=0$ and D3-D3, the ground state $|E_0\ran$ is projected out by
GSO projection and $|E_1\ran$ is kept. However, our D3-$\overline{\rm D3}$
system has opposite GSO projection, and the state with $E_1$ is projected
out and $|E_0\ran$ is kept, leading to a tachyonic state. This state
becomes massless for $\nu=1$ and hence for $E=1$ the system is
tachyon free and stable.

Let us consider what is the spectrum at lower levels. The ground state
is denoted by
\bea
\left| -\frac{1}{2}\nu\right\ran .
\eea
Corresponding to the mode oscillators $c_{n+\nu}$ and $d_{-n-\nu}$,
we have fermionic oscillators $\psi_{n+\nu-\frac12}$ and
$\bar\psi_{-n-\nu-\frac12}$. For $\nu \geq \frac12$,
$\bar\psi_{-\nu+\frac12}, \psi_{\nu-\frac32}$ and transverse oscillators
$\psi^i_{-\frac12} (i=3,\ldots,8)$ give lower states (which remain after
GSO projection)
\bea
\left| \frac{1}{2}(\nu-1) \right\ran &\equiv&
\bar \psi_{-\nu+\frac12} \left| -\frac{1}{2}\nu \right\ran ,\nn
\left| \frac{3}{2}(1-\nu) \right\ran &\equiv&
\psi_{\nu-\frac32} \left| -\frac{1}{2}\nu \right\ran ,\nn
\left| \frac{1}{2}(1-\nu) \right\ran^i &\equiv&
\psi^i_{-\frac12} \left| -\frac{1}{2}\nu \right\ran .
\eea
For $\nu<1$, the first state is the ground state discussed above and
gives tachyonic one. All these states give 8 massless for $\nu=1$,
and the level is degenerate with Ramond sector, in accordance with the
restoration of supersymmetry. Our discussions are quite analogous to those
in Ref.~\cite{BO}. The system has 1/4 supersymmetry in the limit $E\to 1$.
Our result of 8 massless bosonic states is
consistent with this claim because the representation of 8 supercharges
contains $2^4$ states in total, half of which are bosonic.

\section{NCSYM limit of  D3-$\overline{\rm {\bf D3}}$ branes with ${\bf
E}\cdot{\bf B}\neq 0$}

Here we would like to compute the metric for the noncommutative worldvolume
theory on the D3-brane. For this, we shall follow the procedure described
in Refs.~\cite{karch,seiberg}.
We begin with the closed string metric of the diagonal form
$g_{\mu\nu}= {\rm diag}(-|g_{t}|, g_{x}, g_{y},g_{z})$ and
$B_{\mu\nu}$ of the form given in (\ref{BBB}),
which is nothing but  $B_{\mu\nu}$ for the D3-brane.
With this metric and after restoring $2\pi\alpha'$ factors,
the condition for supersymmetry or the ``criticality''
becomes
\begin{eqnarray}
|g_{t}| g_{y}g_{z}= g_{y} (2\pi\alpha' E)^2\,,
\label{electric}
\end{eqnarray}
and hence, $\lambda E=\pm \sqrt{|g_{t}|g_{z}}$
with $\lambda=2\pi\alpha'$.
The open string metric can be identified using the
following relation to the closed string metric~\cite{seiberg}:
\begin{eqnarray}
{1\over g+\lambda B}= {\theta\over\lambda} + {1\over G+\lambda \Phi},
\label{seiberg}
\end{eqnarray}
where  $G$ and $\theta^{\mu\nu}$ are
the open string metric and
the noncommutativity of the worldvolume theory, respectively.
The two form $\Phi$ is free to
choose but there is a natural one for the matrix theory
description~\cite{seiberg}; in the matrix theory
they are given by the relation
$[\hat{\partial}_i, \hat{\partial}_j]=-i \Phi_{ij}$
where $\hat{\partial}_i=-i B_{ij} x^j$ and $[x^i,x^j]=i
\theta^{ij}=i (B^{-1})^{ij}$.

With the value $E$ in (\ref{electric}),
\begin{eqnarray}\label{g+B}
{1\over g+\lambda B}
=-{1\over |g_{t}|g_x (\lambda B)^2 }\left(
\begin{array}{cccc}
g_{x}(\lambda B)^2 + g_z K &  -\lambda B \lambda \tilde{B} \lambda E &
-g_{x} \lambda E \lambda B  & -\lambda E K\\
\lambda B \lambda \tilde{B} \lambda E &  -|g_{t}|(\lambda B)^2 & 0
& -|g_t|\lambda B \lambda \tilde{B}\\
-g_{x} \lambda E \lambda B  & 0  & 0 & |g_t| g_x \lambda B\\
\lambda E K & -|g_t|\lambda B \lambda \tilde{B} & -|g_t| g_x \lambda B &
-|g_t|K
\end{array}
\right)
\end{eqnarray}
where $K= g_x g_y + (\lambda \tilde{B})^2$. Then the Seiberg-Witten limit,
resulting in a noncommutative gauge theory on branes, corresponds to
\begin{eqnarray}\label{SWlimit}
\lambda \sim \sqrt{\epsilon}\,,\ \ \
g_{y}\sim g_{z}\sim \epsilon\ , \ \ \ \epsilon\to 0\ ,
\end{eqnarray}
keeping $B,\ \tilde{B},\ g_{t}$ and $g_x$ fixed.
{}From (\ref{electric}) it is seen that in this limit $E$ is also fixed.
In this limit, one finds that
\begin{eqnarray}
\theta=\left(
\begin{array}{cccc}
0 & 0 & 0 & 0 \\
0 & 0 & 0 & 0 \\
0 & 0 & 0 & -1/B \\
0 & 0 & 1/B & 0
\end{array}
\right)\,,
\end{eqnarray}
for any fixed $\Phi$; this agrees with $\theta$ appearing in the
matrix model description.
The two form for the matrix model is identified as
\begin{eqnarray}
\Phi=\left(
\begin{array}{cccc}
0 & 0 & 0 & 0 \\
0 & 0 & 0 & 0 \\
0 & 0 & 0 & -B \\
0 & 0 & B & 0
\end{array}
\right)\,.
\end{eqnarray}
{}From this and using (\ref{seiberg}), one may
compute $G_{\mu\nu}$ without taking the Seiberg-Witten limit:
\begin{eqnarray}
G_{\mu\nu}=\left((g+\lambda B)^{-1}- {\theta\over\lambda}\right)^{-1}_{\mu\nu}
-\lambda \Phi_{\mu\nu}\,.
\end{eqnarray}
A straightforward evaluation yields
\begin{eqnarray}
G_{\mu\nu}=
\left(
\begin{array}{cccc}
0 & 0 & \lambda B \lambda E/g_z & 0 \\
0 &  K/g_y & 0 & -\lambda B \lambda \tilde{B}/g_y  \\
\lambda B \lambda E/g_z & 0 &  (\lambda B)^2/g_z & 0 \\
0 & -\lambda B \lambda \tilde{B}/g_y & 0 &  (\lambda B)^2/g_y
\end{array}
\right)\,,
\end{eqnarray}
and
\begin{eqnarray}
G^{\mu\nu}=
\left(
\begin{array}{cccc}
-{1\over |g_{t}|} & 0 &  {g_z\over \lambda E \lambda B} & 0 \\
0 & {1\over g_{x}} & 0 & {\lambda \tilde{B}\over g_x \lambda B} \\
 {g_z\over \lambda E \lambda B}  & 0 & 0 & 0\\
0 &  {\lambda \tilde{B}\over g_x \lambda B} & 0 & {g_x g_y +
(\lambda \tilde{B})^2 \over g_x (\lambda B)^2}
\end{array}
\right)\,.
\end{eqnarray}
We note that consistently (using coordinate transformations) we can set
$|g_t|=g_x=1$, and also ${g_y\over \lambda^2}=B^2-{\tilde B}^2$, then
${g_z\over \lambda^2}=E^2$.
$G^{\mu\nu}$ is the metric which appears in the noncommutative action
derived from the background solution. In the background solution,
we take a gauge where the solution is independent of the time and
the D-string is along $\sigma=x^1$ coordinate.
Here we do not have to take  the Seiberg-Witten limit,
but the same metric follows from the limit as mentioned before.

Now let us study the behavior of the open string coupling, $G_s$, in the
decoupling limit. The open string coupling in terms of the closed string
coupling, $g_s$ is given by \cite{seiberg}
\begin{eqnarray}
G_s &=& g_s\ \sqrt{{{\rm det}{(G+\lambda\Phi)}\over {\rm det}{(g+\lambda
B)}}} \cr &=& g_s\ \sqrt{{{\rm det}{(g+\lambda B)}\over {\rm det}\ g}}\ .
\end{eqnarray}
Note that the second line of the above equation is true for any choice of
$\Phi$. Then, imposing the criticality condition (\ref{electric})
\[
{\rm det}{(g+\lambda B)}=-|g_t|g_x (\lambda B)^2\ ,\ \ \
{\rm det}\ g=-|g_t|g_x g_y g_z\ ,
\]
and hence,
\begin{equation}\label{Opencouple}
G_s=g_s\ \sqrt{{(\lambda B)^2\over g_y g_z}}\ .
\end{equation}
In the Seiberg-Witten decoupling limit (\ref{SWlimit}), $G_s \sim g_s
\epsilon^{-1/2}$. Therefore, in order to have a finite open string
coupling, the closed string coupling $g_s$  should behave as
\begin{equation}\label{SWlimit2}
g_s\sim \sqrt{\epsilon}\ .
\end{equation}
In other words in the decoupling limit of (\ref{SWlimit}) and
(\ref{SWlimit2}) the closed strings are essentially
decoupled while the open strings, which are of finite coupling, govern the
dynamics of the system.

We would like to comment that unlike the usual belief that for
${\bf E}\cdot {\bf B} \neq 0$ the Seiberg-Witten decoupling limit makes
the Born-Infeld action imaginary and unstable, is not necessarily true.
The argument of Ref.~\cite{aharony} does not work for the case here
because it is based on the uniform scaling of $g_{\mu\nu}$. The above
computation explicitly demonstrates that there may be a decoupling limit of
noncommutative worldvolume gauge theories even for
${\bf E}\cdot {\bf B} \neq 0$.

\section{NCOS limit}

It is known that strings living in an electric field background in a
particular limit decouples from the closed strings \cite{NCOS}.
In this limit the electric field is taken to some ``critical limit''
\footnote{Note that this criticality in principle is different from ours,
Eq.(\ref{electric}).} and consequently the closed strings mass scale goes
to infinity, while open strings remain of finite mass. In this limit
the closed string coupling, $g_s$ also goes to infinity \cite{NCOS}.
As we will show, although in our case the electric field is already
fixed by the supersymmetry and BPS conditions, still
our D3-$\overline{\rm D3}$ system admits an NCOS limit.

To show that such a limit where the closed strings decouple exists,
let us consider the following limit:
\begin{eqnarray}\label{NCOSlimit}
&& g_t \sim g_z \sim \lambda E \sim \lambda^a \nonumber\\
&& g_x\sim g_y \sim \lambda\,,\ \  B,\ \tilde{B}= {\rm fixed}\ ,
\lambda=2\pi \alpha'\to 0\ .
\end{eqnarray}
with $a < 1$.
The closed string degrees decouple because their mass scale
\begin{eqnarray}
g_t/\alpha' \sim \lambda^{a-1}
\end{eqnarray}
goes to infinity as $\lambda \rightarrow 0$.

In this scaling limit, the open string metric can be read from
Eq.~(\ref{g+B}): \footnote{In this section we set ${\Phi=0}$, and hence
$G^{\mu\nu}$ and $\theta^{\mu\nu}$ are just the symmetric and anti-symmetric
parts of $(g+\lambda B)^{-1}$ respectively.}
\begin{eqnarray}\label{NCOSmetric}
G^{\mu\nu}
={1\over g_x}\left(
\begin{array}{cccc}
 - uv & 0  &
w\sqrt{u}   & 0\\
0  &  1 & 0
& \tilde{B}/B \\
w\sqrt{u}  & 0  & 0 & 0\\
0 &  \tilde{B}/B  & 0 & v
\end{array}
\right)
\end{eqnarray}
where we have defined
\begin{eqnarray}
u\equiv g_z/ |g_t|\,,\ \ v\equiv K/(\lambda B)^2\,,\ \
w\equiv  g_x/ (\lambda B)\,.
\end{eqnarray}
It is clear that $\alpha' G_t^{-1}$ remains finite in the
scaling limit. Hence all the massive modes of open strings
are of finite mass and hence limit (\ref{NCOSlimit}) may lead to an NCOS
theory. We would like to note that although one of the diagonal elements
of the metric (\ref{NCOSmetric}) is zero,
\begin{eqnarray}
{\rm det} G_{\mu\nu}= - {g_x |g_t|\over g_y g_z} (\lambda B)^4\, ,
\end{eqnarray}
and hence metric is not singular. The noncommutativity parameter
\begin{eqnarray}\label{NCOStheta}
\theta^{\mu\nu}
={\lambda\over g_x}\left(
\begin{array}{cccc}
 0 &  \sqrt{u} \tilde{B}/B & 0  &
 v\sqrt{u} \\
- \sqrt{u} \tilde{B}/B &  0  & 0
& 0  \\
0 & 0  & 0 &- w\\
- v\sqrt{u} &  0  & w & 0
\end{array}
\right)
\end{eqnarray}
remains finite and also includes the spacetime component.
As we see
\begin{eqnarray}\label{Openscale}
\lambda_{eff}={\lambda\over g_x},
\end{eqnarray}
is the open string mass scale, as well as the noncommutativity scale.
Without loss of generality, one may consistently set $u=v=1$.

To complete our arguments, we still need to show that in the NCOS limit
(\ref{NCOSlimit}) the open string coupling is finite and also the
{\it massless} closed string states have vanishing coupling. The open
string coupling (\ref{Opencouple}) in the above NCOS limit behaves as
\[
G_s\sim g_s \lambda^{{1-a\over 2}}\ .
\]
Therefore to have a finite $G_s$ the closed string coupling should
behave as
\begin{equation}\label{closedcouple}
g_s \sim \lambda^{-{1-a\over 2}} \to \infty\ .
\end{equation}
Similarly to the usual NCOS cases \cite{NCOS}, our NCOS theory
is actually defined in the strongly coupled closed strings regime.
However, one should note that the coupling for the massless closed string
(supergravity) modes, the ten-dimensional Newton constant $G_N$, differs
from $g^2$ in factors coming from $\sqrt{{\rm det}\ g_{10}}$. In order
to work out the effective coupling for this massless modes, we first need to
fix the scaling of the components of the closed string metric  transverse
to the D3-$\overline{\rm D3}$ system. Let
\begin{equation}
g_{ab}=g_{\perp}\delta_{ab}\ , \ \ \ a,b=1,2\cdots ,6.
\end{equation}
Then the scaling behavior of $g_{\perp}$ can be fixed demanding
that the D3-brane is still carrying one unit of RR-charge. This implies
that
\[ g_{\perp}\sim \lambda^{{1+a\over 2}}\ . \]
On the other hand the effective coupling for the gravitons polarized, say
in $t$-$z$ directions would be
\begin{eqnarray}
g_tG_N&=& {g^2_s\lambda^4\over g_{\perp}^3}\sqrt{{g_t\over g_xg_yg_z}} \cr
      &\sim & \lambda^{{1-a\over 2}}\to 0\ .
\end{eqnarray}
That is, in the decoupling limit defined through (\ref{NCOSlimit}),
(\ref{closedcouple}), the massless closed strings will also
decouple.\footnote{It is straightforward to verify that the effective coupling
for the other graviton polarizations as well as the other supergravity modes
will go to zero as some power of $\lambda^{{1-a\over 2}}$.} Therefore, all
the closed strings, massless and massive, are decoupled and our theory is
completely described by the open strings with mass scale $\lambda_{eff}$
and the coupling $G_s$ given in Eq.~(\ref{Opencouple}).

The above analysis demonstrates that there exists an NCOS limit
related to the supersymmetric D3-$\overline{\rm D3}$.
Note that, in the above computation, $\tilde{B}$ does not play an
essential role and may be set to zero. This in particular implies that
upon a T-duality in the $x^1$ direction our NCOS theory is mapped into another
NCOS theory coming from the  supersymmetric D2-$\overline{\rm D2}$
system of Ref.~\cite{karch,BO}.

\section{Discussion}

In this paper we have studied various generalizations of the supertubes and
brane-antibrane systems. We have shown that there is a 1/4 BPS
D3-$\overline{\rm D3}$ configuration in which ${\bf E}\cdot {\bf B}\neq 0$,
where ${\bf E}$ and ${\bf B}$ are the electric and magnetic background fields
and therefore if we choose ${\bf E}$ to be along the $z$ direction, the
system is specified with two components of the ${\bf B}$ field, one parallel
to {\bf E} and the other one transverse to ${\bf E}$. The latter may be
chosen to be in the $x$ direction. Analyzing the system through the
Born-Infeld action and also the matrix model we showed that in general the
$B$-field component parallel to ${\bf E}$
can be an arbitrary function of $x$ and $y$ while the other component
can be an arbitrary function of $y$. Working out the spectrum of the open
strings attached to D3 and $\overline{\rm D3}$ branes we showed that in
the $E\to 1$ limit tachyon disappears from the spectrum, which is in
agreement with our earlier discussions.
Here we mainly focused on the flat D3-$\overline{\rm D3}$ system. However
similarly to the supertubes case one may think of generalizations with other
cross sectional shapes. Although we have discussed the
D3-$\overline{\rm D3}$ case explicitly, it is also straightforward to
generalize our discussions to the higher-dimensional brane-antibrane systems,
the simplest of which can be obtained through T-dualities.

The supergravity solution for the supertubes has also been worked out
\cite{Emparan}. The similar problem can be  studied for the
general D3-$\overline{\rm D3}$ system we have introduced here.

Here we also briefly discussed the systems with lower supersymmetries.
In particular, we studied the 1/8 BPS solutions in the context of matrix
model. These solutions are matrix version of the noncommutative
tube-D4 system studied in \cite{peet}. Another possible
generalization is on supertubes or D2-$\overline{\rm D2}$ connecting many
D4 branes separated in the transverse space \cite{kklee}, which appear
as  electrically charged  magnetic strings from the view point of
the D4 worldvolume. In the same lines,
generalizing the arguments of section 4, one may construct 6 or 8
dimensional BPS solutions \cite{park} corresponding to D0-D6 or D0-D8
systems. Such solutions can be 1/16, 1/8 or 3/16 BPS.
However, the general classification of the BPS
equations involving electric components is a complicated issue and
may be studied in an independent work.

As another interesting direction to pursue, one may study the behavior of
the  D3-$\overline{\rm D3}$ system under S-duality of type IIB strings.
Being BPS and since S-duality generally does not change the number of
conserved supercharges, we expect  under S-duality D3-$\overline{\rm D3}$
system to be mapped into another stable 1/4 BPS configuration.
In fact using the results of Ref.~\cite{green}, one can show that under the
S-duality transformation $g_s\to {1\over g_s}$, the BPS condition $E=1$,
is again mapped into another similar BPS condition. To be precise,
the background $B$-field of (\ref{BBB}) after the S-duality becomes
\[
B_{01}={\rm sign}(B)\ , \ \ \  B_{12}={1+{\tilde B}^2\over |B|}\ ,
\ \ \  B_{23}= \tilde{B}\, {\rm sign}(B)\ ,
\]
and all the other components zero.
As we see again the electric field, which is now in $x^1$ direction,
squares to one.\footnote{We would like to note that the above expressions
are for the S-dual of the $B$ field on the brane, for the antibrane they are
$B_{01}={\rm sign}(B)\ , \  B_{12}=-{1+{\tilde B}^2\over |B|}$ and
$B_{23}= \tilde{B}\, {\rm sign}(B)$.}
It is then, straightforward to show that for a general $SL(2,Z)$ S-duality
transformation we have similar results; however, we may have non-zero
$B_{23}$ or $B_{13}$ fields as well. Therefore, the set of  general ${\bf
E}\cdot {\bf B}\neq 0$ supersymmetric D3-$\overline{\rm D3}$ systems
is invariant under $SL(2,Z)$ S-duality. Then, one may take the NCSYM or
NCOS limits. We expect that generically these 1/4 BPS NCOS theories
also enjoy the $SL(2,Z)$ symmetry, just similar to the usual 1/2 BPS
NCOS theories~\cite{Russo,CaiO}. However, for some particular values of the
background fields, we expect the NCOS theory to be S-dual to an
NCSYM~\cite{CaiO,lurstwo}. The $SL(2,Z)$ behavior of the BPS
D3-$\overline{\rm D3}$ system deserves a more detailed study which we
postpone to future works.


\noindent{\large\bf Acknowledgments}:

We would like to thank K. Lee, T. Yanagida and P. Yi for enlightening
discussions. The work of D.B. was supported in part by
the Korea Research Foundation Grant KRF-2002-013-D00028.
That of N.O. was supported in part by a Grant-in-Aid for Scientific
Research No. 12640270.
M.M.Sh-J. was supported in part by NSF grant PHYS-9870115 and in
part by funds from the Stanford Institute for Theoretical Physics.

\end{document}